\def\apjl{{\it ApJL}}

\def\mnras{{\it MNRAS}}
\def\apj{{\it ApJ}}

\documentclass{iau} 
\usepackage{graphicx}

\title[Stellar halo profiles and accretion histories in $L_*$ galaxies] 
{Exploring the connection between stellar halo profiles and accretion histories in $L_*$ galaxies}

\author[Nicola C. Amorisco]   
{Nicola C. Amorisco$^{1,2}$
}

\affiliation{$^{1}$Institute for Theory and Computation, Harvard University, 60 Garden Street, Cambridge, MA 02138, USA \\ email: {\tt nicola.amorisco@cfa.harvard.edu} \\[\affilskip]
$^2$Max Planck Institute for Astrophysics, Karl-Schwarzschild-Strasse 1, 85748 Garching, Germany}

\pubyear{2016}
\volume{321}  
\jname{Formation and Evolution of Galaxy Outskirts}
\editors{Eds. A. Gil de Paz, J. C. Lee \& J. H. Knapen}
\begin{document}

\maketitle

\begin{abstract}
I use a library of controlled minor merger N-body simulations, a particle tagging technique and 
Monte Carlo generated $\Lambda$CDM accretion histories to study the highly stochastic process of stellar deposition onto the 
accreted stellar halos (ASHs) of $L_*$ galaxies. I explore the main physical
mechanisms that drive the connection between the accretion history and the density profile of the ASH. 
I find that: i) through dynamical friction, more massive satellites are more effective at delivering their stars deeper into the host;
ii) as a consequence, ASHs feature a negative gradient between radius and the local mass-weighed virial satellite-to-host mass ratio;
iii) in $L_*$ galaxies, most ASHs feature a density profile that steepens towards sharper logarithmic slopes at increasing radii, 
though with significant halo-to-halo scatter; iv) the ASHs with the largest total ex-situ mass are such because of the chance accretion
of a small number of massive satellites (rather than of a large number of low-mass ones).
\end{abstract}

\firstsection 
\section{Introduction}

The accreted stellar halo (ASH) of a galaxy represents a record of the accretion history of the galaxy itself. 
Its assembly is determined by a large number of free parameters, including the structural properties of 
each accreted satellite (virial mass, concentration, stellar content, morphology), the orbital properties of each 
accretion event (energy and angular momentum at infall), the structural properties of the host itself during accretion. 
This implies a significant degree of stochasticity, as shown by the observed halo-to-halo scatter (e.g., van Dokkum et al. 2014) 
and by the dichotomy between the 
`broken' and sharply declining density profile of the stellar halo of the Milky Way (e.g., Deason et al. 2013) and the 
more extended halo of Andromeda, whose density profile is well described by a single power-law (e.g., Gilbert et al. 2012, Ibata et al. 2014).

\section{Individual contributions to the accreted stellar halo}

In Amorisco (2015) I have isolated the main ingredients that shape the contribution of each accreted satellite to the ASH. 
I adopted a simplified approach and assumed that the contributing dwarfs are dark matter dominated (as expected for the case of an $L_*$
host) and ignored the gravitational influence of the host's disk. Combined with the halo mass-concentration relation (e.g., Gao et al. 2008, Ludlow et al. 2014),
this reduces the structural properties of each minor merger to two dimensionless parameters. Two additional parameters
characterise the orbital properties of the satellite at accretion (e.g., Benson 2005, Jiang et al. 2015).
The locus of this parameter space that is relevant to a $\Lambda$CDM cosmology is explored with a library of minor merger 
N-body simulations, in which stars are assigned to the most bound 5$\%$ of the satellite's particles, using a standard 
particle tagging technique (e.g., Bullock \& Johnston 2005, Cooper et al. 2010). This study shows that dynamical friction 
is a major player in shaping stellar deposition, allowing only the most massive (and/or concentrated) accreted satellites to deposit their stars 
in the innermost regions of the host. Orbital radialisation by dynamical friction causes the stellar populations deposited by 
such most massive accretion events (virial satellite-to-host mass ratio at accretion $\gtrsim 1/20$) to bear little memory 
of the details of the orbital properties of the progenitor at infall.   

\begin{figure}[b]
\begin{center}
 \includegraphics[width=\textwidth]{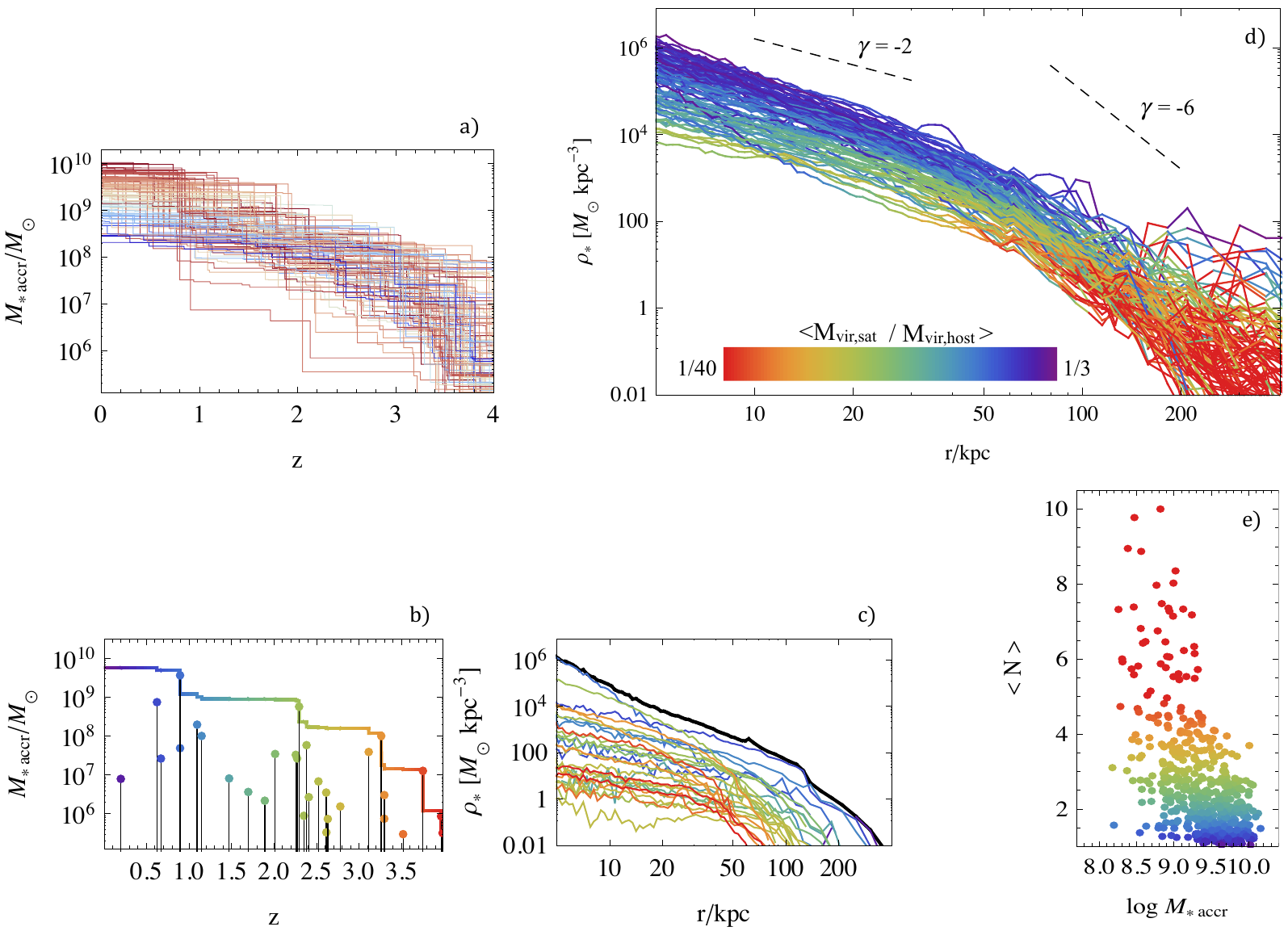} 
 \caption{Panel a): a sample of 150 Monte Carlo generated stellar accretion histories for $L_*$ 
 galaxies sharing a final virial mass of $2\times 10^{12}M_{\odot}$. Panels b) and c): an example for the 
 procedure of assembly of the accreted stellar halo using the combination of a given accretion history and 
 of a library of minor merger simulations; each contribution to the halo is color-coded by accretion redshift.
 Panel c): the spherically averaged density profiles of the 150 accreted stellar haloes corresponding to the 
 accretion histories of panel a), color-coding refers to the local mass-weighed virial satellite-to-host mass.
 Panel e): the correlation between the total accreted stellar mass in the halo and the number of main 
 accretion events (see text for details).}
   \label{fig1}
\end{center}
\end{figure}

\section{Towards an understanding of the variability of ASHs in $L_*$ galaxies}

I use the library just described to assemble 500 ASHs, for galaxies that share a virial mass of
$M_{200}(z=0)=2\times 10^{12}M_{\odot}$. I use Monte Carlo generated accretion 
histories (Fakhouri et al. 2010), 150 of which are displayed in {\it panel a)} of Fig.~1, color-coded by the 
total accreted stellar mass. Satellite stellar masses are assigned based on 
a redshift-independent abundance matching relation (Garrison-Kimmel et al. 2014, 0.3 dex scatter).
{\it Panels b)} and {\it c)} exemplify the assembly of an individual ASH: each accreted and disrupted satellite (color-coded by its accretion redshift)
is associated to the spherically averaged density profile of the stars it deposits in the halo. These are retrieved 
from the library using the relevant time-interval between accretion and $z=0$, and are  
re-scaled to physical quantities according to the dimensional properties of the merger at hand. 

{\it Panel d)} displays the spherically averaged density profiles of 150 ASHs built in this manner. At each radius, 
the halo-to-halo scatter approaches 2 dex, and increases at $r\gtrsim100$, together with an increasing amount of not fully phase-mixed substructure from recent accretion events.
Although with a significant scatter, on average, ASHs share a logarithmic density slope $\gamma\sim-2$ within 20~kpc, and
become steeper with radius, as shown by the dashed guiding lines. The details of 
this steepening are highly variable: some profiles have marked and sharp breaks, others `roll' gently towards steeper and steeper slopes, other
remain comparatively shallower. The radii where such transitions take place are equally variable. The color-coding in {\it panel d)} indicates the local mass-weighted satellite-to-host virial mass ratio. On average, the innermost regions of the ASH are contributed by satellites that have larger virial mass ratios 
at accretion. This gradient has been observed in cosmological hydrodynamical simulations (Rodriguez-Gomez et al. 2016) and I conclude is a direct consequence of dynamical friction (e.g., Amorisco 2015). 
Color-coding in {\it panel d)} reveals that the local mean virial mass ratio also correlates positively with the local density. {\it Panel e)} 
confirms this link by showing a scatter plot of the total accreted stellar mass of the ASH against the `number of main accretion 
events' $\langle N\rangle$ (i.e. the ratio between the total accreted stellar mass of the ASH and the mean stellar mass of the contributing satellites, 
mass-weighted by stellar mass itself). The most massive ASHs result from the accretion of just one/two particularly massive satellites, 
which dominate the ex-situ mass. 

Although this technique represents a highly simplified approach, it allows for an efficient exploration of the significant stochasticity 
of ASHs. Physical ingredients that are neglected here (the host's stellar disk, the morphologies of the 
accreted dwarfs, any post-accretion star formation etc.) will result in even increased variability. 
Detailed analysis of a large sample of ASHs concentrating on the systematic correlations that connect density profile and  
accretion history is the subject of a forthcoming work (Amorisco 2016).

\end{document}